\documentclass[10pt,conference]{IEEEtran}
\usepackage{epsfig,graphicx,subfigure,psfrag,amsmath,cases,bm}
\usepackage{latexsym,amssymb,algorithm,mathtools}
\usepackage{algorithmic}
\usepackage{color}
\usepackage{scrtime}
\usepackage{stfloats}
\usepackage{tablefootnote}
\usepackage{cite}
\usepackage{amsfonts,mathabx}
\usepackage{textcomp}
\usepackage{caption}
\usepackage{xcolor,capt-of}
\usepackage{multirow}
\usepackage{subfigure}
 
\usepackage[left=0.625in,right=0.625in,bottom=0.95in,top=0.73in]{geometry}
\allowdisplaybreaks
\IEEEoverridecommandlockouts

\newcommand\scalemath[2]{\scalebox{#1}{\mbox{\ensuremath{\displaystyle #2}}}}

\def\BibTeX{{\rm B\kern-.05em{\sc i\kern-.025em b}\kern-.08em
T\kern-.1667em\lower.7ex\hbox{E}\kern-.125emX}}
\begin{document}
\title{Energy-Aware Resource Allocation and Trajectory Design for UAV-Enabled ISAC}
\author{\IEEEauthorblockN{Ata Khalili$^{\star}$,~Atefeh Rezaei$^{\dag}$, Dongfang Xu$^{\ddag}$, and Robert Schober$^{\star}$}
\small $^{\star}$Friedrich-Alexander-University Erlangen-Nurnberg, Germany,
$^{\dag}$Technical University of Berlin, Germany,\\$^{\ddag}$The Hong Kong University of Science and Technology, Hong Kong\vspace{-7.5mm}\\
\thanks{This work was supported partly by the Federal Ministry of Education and Research of Germany under the program of “Souveran. Digital. Vernetzt.” joint project 6G-RIC (project identification number: PIN 16KISK023) and also in part by the Deutsche Forschungsgemeinschaft (DFG, German Research Foundation) GRK-2680 – Project-ID 437847244.}}
 \maketitle
\begin{abstract}
In this paper, we investigate joint resource allocation and trajectory design for multi-user multi-target unmanned aerial vehicle (UAV)-enabled integrated sensing and communication (ISAC). To be compatible with practical UAV-based sensing systems, sensing is carried out while the UAV hovers.~In particular, we jointly optimize the two-dimensional trajectory, the velocity, and the downlink information and sensing beamformers of a fixed-altitude UAV for minimization of the average power consumption, while ensuring the quality of service of the communication users and the sensing tasks. To tackle the resulting non-convex mixed integer non-linear program (MINLP), we exploit semidefinite relaxation, the big-M method, and successive convex approximation to develop an alternating optimization-based algorithm.~Our simulation results demonstrate the significant power savings enabled by the proposed scheme compared to two baseline schemes employing heuristic trajectories.
	\end{abstract}
	\vspace{-2mm}
	\section{Introduction}
 	\vspace{-1mm}
 Integrated sensing and communication (ISAC) has lately drawn significant attention as a promising technology to increase the spectrum efficiency and enable the sharing of the physical infrastructure for sensing and communications in sixth-generation (6G) wireless communication systems \cite{ISAC6G}.
In this regard, the authors of \cite{mu-mimo-jsc,jsc-mimo-radar} studied transmit beamforming for ISAC systems, where a least-squares problem was formulated to obtain the ideal beampattern for sensing while guaranteeing a required signal-to-interference-plus-noise ratio (SINR) of the communication users. However, these works considered terrestrial ISAC systems which are typically impaired by surrounding obstacles and scatterers on the ground blocking the line of sight (LoS) to the sensing targets.
	
 On the other hand, unmanned aerial vehicle (UAV)-aided wireless communication has drawn significant attention as a result of its simple deployment and favorable channel characteristics \cite{SurveyUAV,DUAV}. In fact, UAVs can provide LoS links to the ground,  which are also desirable for sensing, as target detection and parameter estimation require LoS links between the sensing transceivers and the sensing targets. Furthermore, due to their high maneuverability, UAVs can significantly reduce the typically high sensing powers as they can approach to the targets\cite{CSUAV}. Despite these promising features, only few works in the existing literature have studied UAV-enabled ISAC \cite{maneuver,UAVISAC1,ThUAVISAC,UAVISACSecure}. The authors in \cite{maneuver} optimized the trajectory, transmit beamforming, and radar signals of a UAV-enabled ISAC system to improve the communication data rate while ensuring a required sensing beampattern gain. In \cite{UAVISAC1}, a periodic sensing and communication scheme for UAV-enabled ISAC systems was introduced and the achievable rate was maximized by jointly optimizing the UAV's trajectory, transmit precoder, and sensing start time subject to sensing frequency and beampattern gain constraints. Besides, in \cite{ThUAVISAC}, user association, sensing time selection, beamforming, and the UAV trajectory were jointly optimized to boost the total achievable data rate of an UAV-based ISAC system. The authors in \cite{UAVISACSecure} proposed a novel integrated sensing, jamming, and communication framework for UAV-enabled downlink communications to maximize the number of securely served users while considering a tracking performance constraint. Yet, the authors of \cite{maneuver,UAVISAC1,ThUAVISAC,UAVISACSecure} focus only on beampattern gain optimization for target sensing and do not take into account the sidelobes of the beams which waste energy and may cause unwanted interference\cite{jsc-mimo-radar,mu-mimo-jsc}. Besides, sensing was performed while the UAV was moving, which may degrade the sensing accuracy \cite{UAVr}. 
In fact, in practical UAV-based sensing systems, the UAV senses only during hovering \cite{UAVr}. Therefore, in this paper, we incorporate this feature into the  problem formulation. This has several benefits. First, during hovering, the effect of UAV jittering is smaller as compared to when the UAV moves which results in a better
sensing performance \cite{Jittering}. Second, when the UAV hovers above the target a predetermined fixed beampattern can be designed i.e., the beampattern does not need to be continuously adjusted based on the UAV's flight path, which reduces the design complexity significantly. Third, hovering during sensing circumvents the UAV-induced Doppler shift, simplifying the sensing data signal processing.   
			In this paper, we optimize the average UAV power consumption taking into account the quality of service (QoS) requirements of the communication users and the sensing tasks.~The main contributions of this paper can be summarized as follows:
	\begin{itemize}
	    \item We investigate the joint resource allocation and trajectory design for an UAV-enabled ISAC system to minimize the average power consumption of the UAV.~To this end, we formulate an optimization problem where also the time when the UAV hovers for sensing is subjected to optimization, which leads to a non-convex mixed integer non-linear program (MINLP).
	    \item We develop an alternative optimization (AO) based resource allocation algorithm to solve the optimization problem. In particular, we obtain a low-complexity sub-optimal solution for the formulated highly non-convex MINLP by exploiting semi-definite relaxation, the big-M method, and successive convex approximation (SCA). 
	    \item Simulation results demonstrate the superiority of the proposed resource allocation algorithm design compared to two baseline schemes in terms of the average power consumption. Besides, we  show that the proposed algorithm also ensures that the UAV hovers above the target during sensing. 
	\end{itemize}
	\textit{Notations:} In this paper, matrices and vectors are denoted by
boldface capital letters $\mathbf{A}$ and lower case letters $\mathbf{a}$, respectively.~$\mathbf{A}^T$,~$\mathbf{A}^H$, $\text{Rank}(\mathbf{A})$, and $\text{Tr}(\mathbf{A})$ are the transpose,~Hermitian conjugate transpose, rank, and trace of matrix $\mathbf{A}$, respectively. 
$\mathbf{A}\succeq\mathbf{0}$ denotes a positive semidefinite matrix. $\mathbf{I}_N$ is the $N$-by-$N$ identity matrix.
The absolute value of a complex scalar and the Euclidean norm of a complex vector are denoted by $|\cdot|$ and $\|\cdot\|$, respectively. 
$\mathcal{C}\mathcal{N}(\boldsymbol{\mu},\,\mathbf{C})$ represents the circularly symmetric complex Gaussian (CSCG) distribution with mean $\boldsymbol{\mu}$ and covariance matrix $\mathbf{C}$.
Finally, $\mathbb{C}^{M\times \!N}$ represents an $M\times~\!\!N$ dimensional complex matrix and $\nabla_{\mathbf{x}}$ is the gradient with respect to~$\mathbf{x}$. 
\begin{figure}
	\centering
	\includegraphics[width=0.95\linewidth]{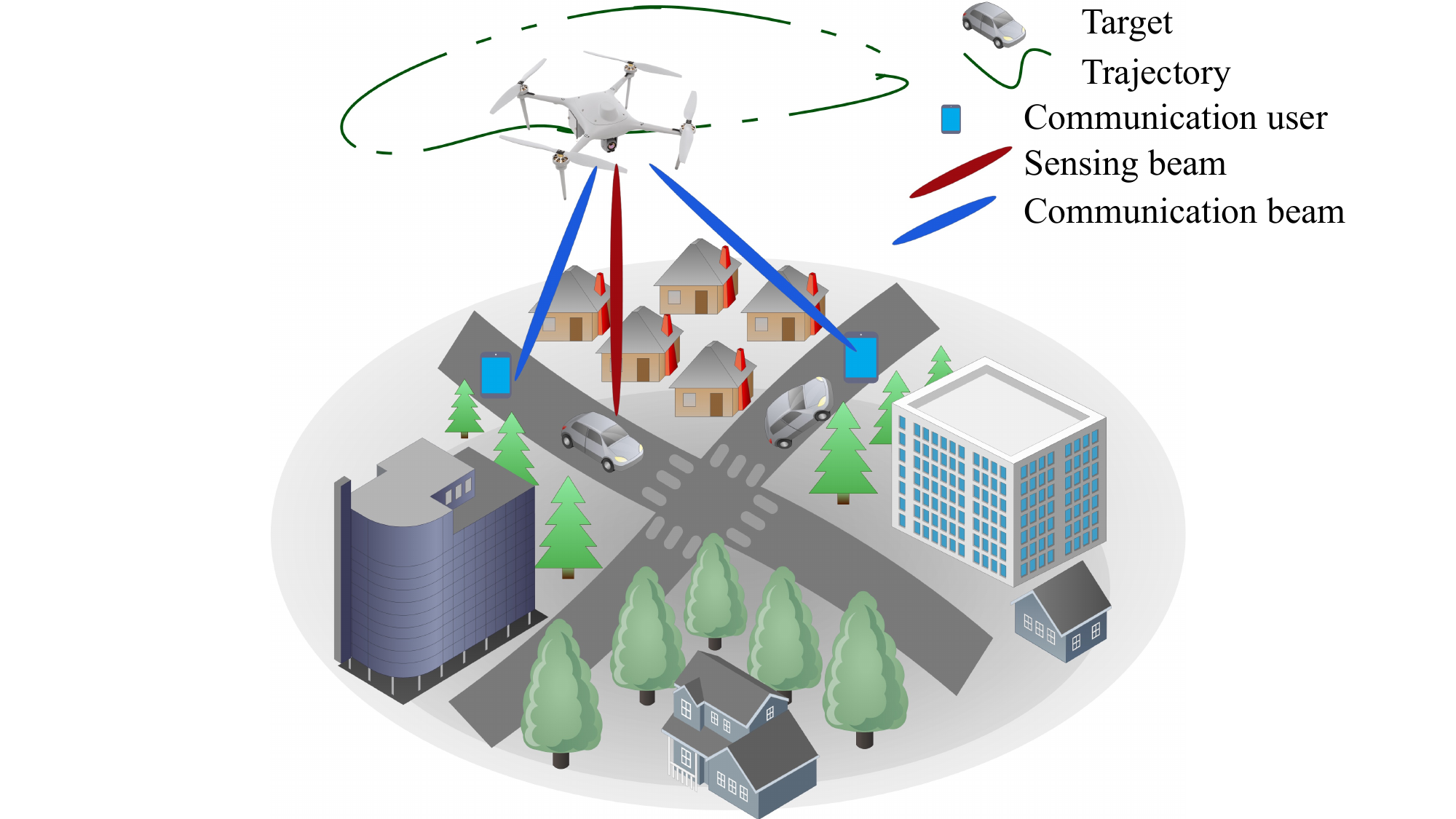}
	\vspace{-2mm}
	\caption{ \small Joint communication and sensing in UAV-assisted network comprising  $E=2$ sensing targets and $K=2$ communication users.   }\label{fig:sys}
 \vspace{-3mm}
	\end{figure}
\vspace{-6mm}
	\section{System Model}
\vspace{0mm}
In this paper, we consider a rotary-wing UAV-assisted ISAC system that provides downlink communication services for $K$ communication users and senses $E$ potential targets as shown in Fig.~1. The UAV's total flying time $ T $ is divided into $ N $ time slots of duration $ \delta_t =\frac{T}{N}$. Each time slot is assumed to be sufficiently small, such that the location of the UAV can be assumed to be approximately constant during a time slot which facilitates the trajectory and beamforming design for ISAC. We adopt a three-dimensional (3D) Cartesian coordinate system where the horizontal location of the UAV and the $k^{\text{th}}$ communication user in time slot $n$ are denoted by ${\mathbf{q}}[n] = {\big[q_x[n],q_y[n]\big]^{T}}$  and ${\mathbf{d}}_{k}= {\big[d_{x_k}, d_{y_k}\big]^{T}}$, respectively.~Moreover, it is assumed that the UAV flies in the $ x-y $ plane at fixed altitude $ H $ subject to air traffic control. The UAV is equipped with a uniform linear array (ULA) with $M$ antennas and transmits simultaneously information signals $c_k[n]$, $c_{k}\sim \mathcal{CN}(0,1)$, $k \in \{1,...,K\}$, to $K$ communication users. Hence, the baseband transmit signal of the UAV in time slot $n$ can be expressed as $\mathbf{x}[n] = \sum_{k=1}^K \mathbf{w}_k[n] c_k[n],$
where $\mathbf{w}_k[n] \in \mathbb{C}^{M \times 1}$ denotes the transmit beamforming vector for user $k$. 

\subsection{ISAC Frame Structure}
\begin{figure}[t]
	 	\centering
	\includegraphics[width=0.95\linewidth]{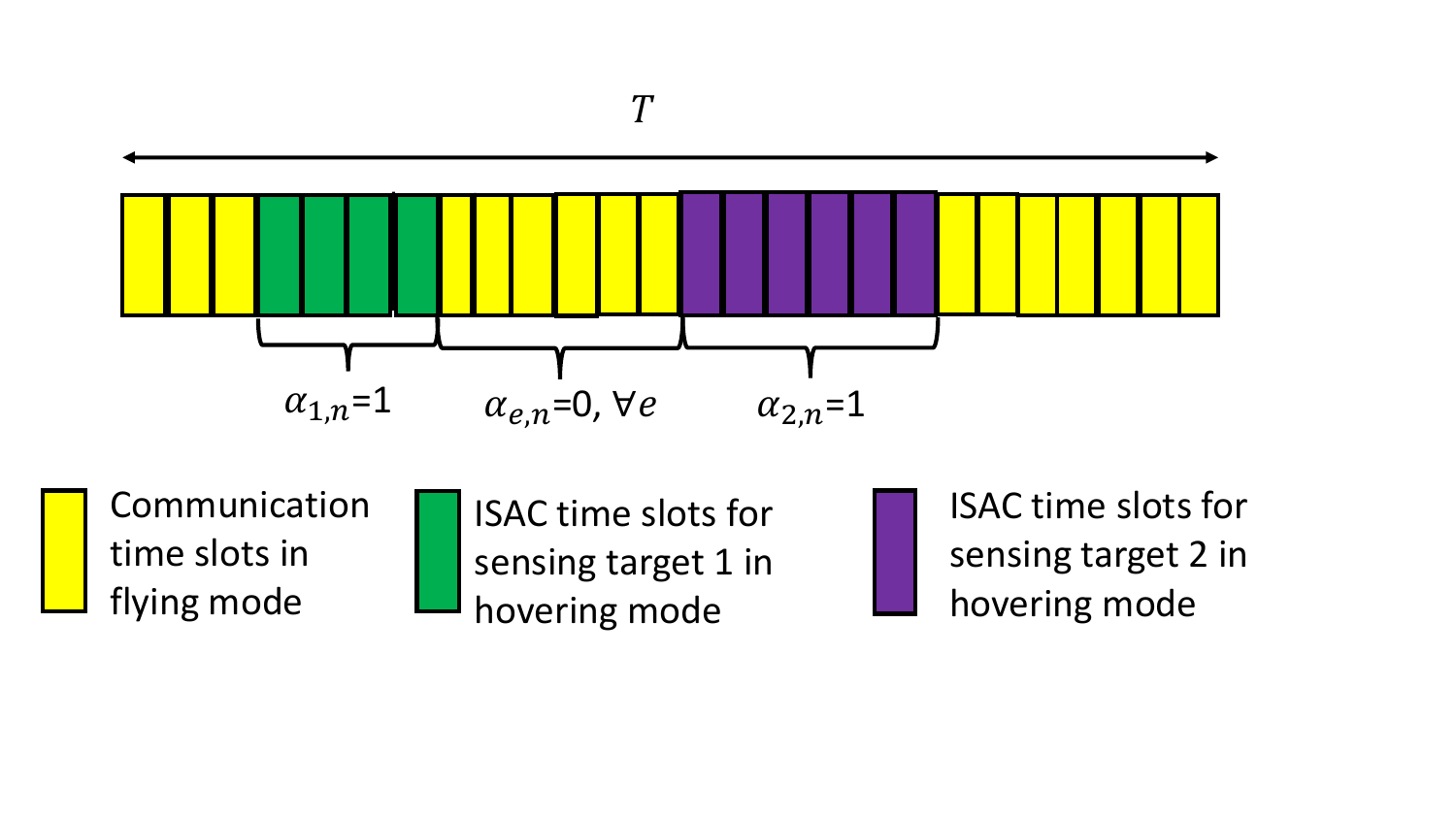}
   \vspace{-10mm}
	 	\caption{ \small Proposed ISAC frame structure where $T$ is the total flying time.}.\label{frame}
   \vspace{-10mm}
	 	\end{figure}

The proposed frame structure for UAV-ISAC is shown in Fig. \ref{frame}. The UAV can communicate with the communication users in all time slots. However, the UAV can use only a maximum of $N_{s}^{\max}$ time slots for sensing. At most one target is sensed in a given time slot to maximize the sensing performance by focusing the transmit beam on the target. However, in which time slots sensing is performed is part of the optimization. To this end, we define $\alpha_{e,n}$ as the sensing indicator for target $e$,  $e \in \{1,...,E\}$. In particular, if $\alpha_{e,n}=1$, target $e$ is sensed in the $n$-th time slot, during which the UAV hovers above the target; otherwise, $\alpha_{e,n}=0$. 
\subsection{Radar and Communication Models}
The location of the potential target on the ground is denoted by ${\mathbf{d}}_{e} = {\big[d_{x_e}, d_{y_e}\big]^{T}} \in {\mathbb{R}^{2 \times 1}}$. The value of ${\mathbf{d}}_{e}$, $e \in \{1,...,E\}$,  is predetermined based on the specific sensing task\footnote{${\mathbf{d}}_{e}$ could be the estimated location of a mobile target for target tracking applications or it could be a fixed location in the region of interest for target detection applications \cite{maneuver,UAVISAC1,ThUAVISAC}.}. The UAV emits a narrow beam towards the direction of the target and extracts the desired sensing information from the received echo signals. We assume the communication signals are also exploited for target sensing. Thus, the transmit beampattern gain from the UAV in the direction of target $e$ is given by $\mathcal{P}(\mathbf{w}_{k},\mathbf{q}[n],\mathbf{d}_{e})=\mathbf{a}^H(\mathbf{q}[n],\mathbf{d}_{e})~ \big(\sum^{K}_{k=1}\mathbf{w}_{k}[n]\mathbf{w}_{k}^{H}[n]\big)~\mathbf{a}(\mathbf{q}[n],\mathbf{d}_{e}) $, where
$\mathbf{a}(\mathbf{q}[n],\mathbf{d}_{e})\hspace{-0.25mm}= \big[1,e^{j 2\pi \frac{\hat{d}}{\lambda} \cos (\theta(\mathbf{q}[n],\mathbf{d}_{e}))},...,$ $e^{j 2\pi\frac{\hat{d}}{\lambda} (M-1) \cos(\theta(\mathbf{q}[n],\mathbf{d}_{e}))}\big]^T$ is the steering vector of the uniform linear array equipped at the UAV, $\theta(\mathbf{q}[n],\mathbf{d}_{e})=\arccos\big(\frac{H}{\sqrt{\|\mathbf{q}[n]-\mathbf{d}_{e}\|^{2}}+H^{2}}\big)$ is the angle of departure corresponding to target $e$, $\lambda$ is the carrier wavelength, and $\hat{d}$ denotes the spacing between two adjacent antennas. 

 
Next, the echo signal received at the UAV in time slot $n$ is given by
$\mathbf{r}_{e}[n]=  {\mathbf{H}_{e}[n]
\bigg(\sum_{k=1}^K \mathbf{w}_k[n] c_k[n]}
\bigg) + \mathbf{z}[n]$,
where $\mathbf{z}\sim\mathcal{C}\mathcal{N}(\mathbf{0},\sigma^{2}_{e}\mathbf{I}_{M})$ is the received additive white Gaussian noise (AWGN) at the UAV and $\mathbf{H}_{\mathrm{e}}[n]$ is the round-trip channel matrix, which is given by
$\mathbf{H}_{e}[\hspace{-0.25mm}n\hspace{-0.25mm}] \hspace{-1mm}=\hspace{-1mm}\frac{\epsilon_{e}\hspace{-0.25mm}[\hspace{-0.25mm}n\hspace{-0.25mm}]\beta_0}{2\Psi_{e}[n]} \mathbf{a}{(\textbf{q}[n],{\textbf{d}}_e)}\mathbf{a}^{H}{(\textbf{q}[n],{\textbf{d}}_e)}$,
where $\beta_{0}$ denotes the channel power gain at the reference distance of $d_{0} = 1$~m and $\Psi_{\mathrm{e}}[n]={{{\sqrt{\left\| {{\mathbf{q}[n]} - {\mathbf{d}_e}} \right\|^2+H^2}}}}$. Moreover, $\epsilon_{e}\hspace{-0.25mm}[\hspace{-0.25mm}n\hspace{-0.25mm}]\hspace{-1mm} =\hspace{-1mm} \sqrt{\hspace{-1mm}\frac{\vartheta_{e}}{4\pi \Psi^2_e[n]}}$ denotes the reflection coefficient of target $e$ in time slot $n$, and $\vartheta_{e}$ is the radar cross-section of target $e$ \cite{Radar}. 
To achieve adequate sensing performance, we require the accumulated sensing SNR of target $e$, i.e., $\Gamma_{e}$, to be higher than a preset minimum threshold $\mathrm{SNR}_{e}^{\mathrm{th}}$, where $\Gamma_{e}$ is given by
\vspace{-1mm}
\begin{align} 
\scalemath{0.95}{\Gamma_{e}\triangleq\sum_{n=1}^{N} \frac{\alpha_{e,n}\vartheta_{e}\beta_{0}^{2}\mathbf{a}^H(\mathbf{q}[n],\mathbf{d}_{e})\bigg(\overset{K}{\underset{k=1}{\sum }}\mathbf{w}_k[n] \mathbf{w}^H_k[n]\bigg)\mathbf{a}(\mathbf{q}[n],\mathbf{d}_{e})}{16\pi \Psi^4_{e}[n]\sigma^{2}_{e}}},
\end{align}
where $\text{SNR}^{\text{th}}_{e}$ is the minimum SNR required at the UAV for sensing target $e$.
 The channel vector between the UAV and user $k$ is denoted by $\mathbf{h}_{k}$, and given by
	${\mathbf{h}_{k}}{[n]} = \frac{{{\beta _0\mathbf{a}{(\textbf{q}[n],{\textbf{d}}_k[n])}}}}{{{{\sqrt{\left\| {{\mathbf{q}[n]} - {\mathbf{d}_k[n]}} \right\|^2+H^2}}}}}$, based on the free space channel model.
Then, the received signal at user $k$ can be written as
	   $  y_{k}[n]=\mathbf{h}^{H}_k[n]\bigg(\sum_{k=1}^K \mathbf{w}_k[n] c_k[n] \bigg)+z_{k}[n]$,
	 where $n_{k}\sim\mathcal{CN}(0,\sigma^{2}_{k})$ is the AWGN at user $k$.~Consequently, the received SINR of user $k$ in time slot $n$ is given by
     \vspace*{-2mm}
		\begin{equation}\label{sinr}
	{\gamma_{k}}{[n]}= \dfrac{\big|\mathbf{h}^{H}_k[n]\mathbf{w}_k[n]\big|^2}{\sum_{i\neq k}^{}\big|\mathbf{h}^H_k[n]\mathbf{w}_{i}[n]\big|^2+\sigma^2_k}.
		\end{equation}
\subsection{Power Consumption Model}
The propulsion power consumption depends on the flying mode of the UAV \cite{DUAV,Rotary}. In particular, the aerodynamic power consumption for rotary-wing UAVs is a function of its flight velocity $\mathbf{v}[n]\in \mathcal{R}^{2\times 1}$\cite{Rotary}. The total UAV power consumption in time slot $n$ during hovering and flight is given by
$P(\mathbf{v}[n]) = \sum_{e=1}^{E}\alpha_{e,n}P_{\text{hover}}[n]+(1-\sum_{e=1}^{E}\alpha_{e,n})P_{\text{fly}}(\mathbf{v}[n]) \label{eqn:P_flight_I}$\cite{Rotary},
where $P_{\text{hover}}[n]=P_o+P_i$ and $P_{\text{fly}}(\mathbf{v}[n]) $=$
P_o \bigg(\frac{3 \|\mathbf{v}[n]\|^2 }{\Omega^2 r^2} \bigg) + P_i  \bigg[\left( \sqrt{1+\frac{\|\mathbf{v}[n]\|^4}{4 v_0^4}}-\frac{\|\mathbf{v}[n]\|^2}{2v_0^2}\right)^{1/2}-1\bigg]+ \frac{1}{2} r_0 \rho s A_{\mathrm r} \|\mathbf{v}[n]\|^3$, respectively.
The parameters of the adopted power consumption model are summarized in Table \ref{notations}\cite{Rotary}.
 \begin{table}[t]
\centering
\scriptsize
  \caption{Parameters in the power consumption model\cite{Rotary}.}\label{notations} 
\begin{tabular}{ c | c }
  \hline			
  Notations & Definitions \\ \hline
  $\Omega=300$ & Blade angular velocity in radians/second \\
  $r=0.4$ & Rotor radius in meter \\
  $\rho=1.225$ & Air density in $\mathrm{kg/m^3}$ \\
  $s=0.05$ & Rotor solidity in $\mathrm{m^3}$ \\
  $A_{\mathrm r}=0.503$ & Rotor disc area in $\mathrm{m^2}$ \\
  $P_o=80$ & Blade profile power during hovering in Watt\\
  $P_i=88.6$ & Induced power during hovering in Watt\\
  $v_0=4.03$ & Mean rotor induced velocity in forward flight in m/s\\
  $r_0=0.6$ & Fuselage drag ratio \\
  \hline
\end{tabular}
\vspace*{-6mm}
\end{table}
\section{Problem Formulation}
\vspace{-1mm}
In this paper, we aim to minimize the average power consumption of the UAV by jointly optimizing the beamforming for information transmission and sensing, the time slots when the UAV hovers above the target for sensing, $\alpha_{e,n}$, the UAV trajectory ($ \mathbf{q} $), and the velocity of the UAV ($ \mathbf{v} $),  while guaranteeing the QoS of the communication users and the sensing targets.~As a result, the optimization problem is mathematically formulated as follows:
\vspace{-2mm}
\begin{align}
& \mathcal{P}_{1}: \mathop {{\rm{min}}} \limits_{\scriptstyle{\boldsymbol{\Xi}}}\mathcal{O}bj\triangleq\frac{1}{N}\sum_{n=1}^{N}\bigg(\sum_{k=1}^{K} \|\mathbf{w}_k[n]\|^2 +\nonumber\\&\sum_{e=1}^{E}\alpha_{e,n}P_{\text{hover}}[n]+(1-\sum_{e=1}^{E}\alpha_{e,n})P_{\text{fly}}(\mathbf{v}[n])\bigg)\nonumber\\
	\text{s.t.}~~
	&\text{C}1:\scalemath{0.9}{\sum_{k=1}^{K} \|\mathbf{w}_k[n]\|^2\leq P_{\max}},\nonumber\\
	&\text{C}2:\scalemath{0.9}{\frac{1}{N}\sum_{n=1}^{N}\log_2(1+\gamma_{k}[n])\geq R_{\min}^{k}, \forall k},\nonumber\\
	&\text{C}3:\scalemath{0.9}{\alpha_{e,n}\bigg\|\sum_{k=1}^{K}\mathbf{w}_k[n] \mathbf{w}^H_k[n] -\mathbf{R}_{d}\bigg\|_{F}^2\!\!\!\leq \epsilon},~\text{C}4:\scalemath{0.9}{\Gamma_{e}\geq\text{SNR}_{\text{e}}^{\text{th}}},\nonumber\\
&\text{C}5:\sum_{e=1}^{E}\alpha_{e,n}\leq 1, \forall n,~\text{C}6:\sum_{n=1}^{N}\alpha_{e,n}\leq N_{s}^{\max},\forall e,\nonumber\\
&\text{C}7: \sum_{e=1}^{E}\alpha_{e,n}\big\|\mathbf{q}[n]-\mathbf{d}_{e}\big\|^{2}\leq D,\nonumber\\
&\text{C}8:\mathbf{q}[n+1]=\mathbf{q}[n]+(1-\sum_{e=1}^{E}\alpha_{e,n})\mathbf{v}[n]\delta_t,\forall n,e,\nonumber\\ 
&\text{C}9:\big\|\mathbf{v}[n+1]-\mathbf{v}[n]\big\|\leq a_{\max}\delta_t,\forall n,\nonumber\\
& \text{C}10:\big \|\mathbf{v}[n]\big \| \leq (1
	-\sum_{e=1}^{E}\alpha_{e,n})v_{\max},\forall n,\nonumber\\
	&\text{C}11:\alpha_{e,n} \in \{0,1\}, \forall e,n \label{ie}.
\end{align}
In \eqref{ie}, $\boldsymbol{\Xi}=\{ \mathbf{w}_{k}[n], \mathbf{q}[n],\mathbf{v}[n],\alpha_{e,n}\}$ is the set of optimization variables.
\text{C}1 limits the transmit power of the UAV, where $P_{\max}$ is the maximum transmit power. \text{C}2 guarantees that the average achievable data rate of the communication users does not fall below the minimum data rate $R^{k}_{\min}$. \text{C}3 ensures that the difference between the desired radar beampattern and the actual beampattern of the transmitted signal does not exceed a predefined threshold $\epsilon$. In particular, the predesigned highly-directional sensing beampattern is characterized by the covariance matrix of the desired waveform, i.e., $\mathbf{R}_{d}$ \footnote{This constraint can be used to synthesize a focused beam with small sidelobes for sensing minimizing interference and clutter. Note that $\mathbf{R}_{d}$ is independent of the trajectory as $\text{C}7$ ensures the UAV always hovers above the target for sensing.}. \text{C}4 ensures the accumulated SNR of the reflected signal at the UAV does not fall below a threshold. \text{C}5 indicates that at most one target can be sensed in a time slot. \text{C}6 limits the maximum number of time slots for sensing to $N_{s}^{\text{max}} $.
\text{C}7 ensures the horizontal distance between the UAV and the target is smaller than $D$. For small $D$, the UAV will hover above the target during sensing
. $\text{C}8$ models the evolution of the trajectory of the UAV based on its flight velocity. Furthermore, 
	 \text{C}9 and \text{C}10 limit the maximum acceleration and velocity of the UAV to $a_{\text{max}} $ and $ v_{\text{max}} $, respectively. Finally, \text{C}11 ensures that the sensing indicator is an integer variable. 
	\vspace{-1mm}
	\section{Solution of the optimization problem}
Optimization problem $\mathcal{P}_{1}$ is non-convex due to the coupling between the optimization variables and the non-convexity of constraints $\text{C2}-\text{C4},\text{C7}, \text{C8}$,~$\text{C10}$,~$\text{C11}$, and the objective function. In general, it is very challenging to find a globally optimal solution to the non-convex optimization problem. Therefore, we propose an iterative algorithm based on the AO approach to obtain a low-complexity suboptimal solution. In particular, 
we first optimize the beamforming matrices and the sensing indicator, and then we jointly optimize the trajectory and velocity of the UAV.
	\vspace{-1mm}
	\subsection{Beamforming and Sensing Indicator Optimization}
	First, we assume that the position and velocity of the UAV are fixed and we aim to optimize the beamformers for communication and sensing. To do so, we employ semidefinite programming (SDP) and define $\mathbf{W}_{k}=\mathbf{w}_{k}\mathbf{w}_{k}^{H} $, where $\mathbf{W}_{k}\succeq 0$ and $\text{Rank}(\mathbf{W}_{k})\leq 1$. One obstacle for solving optimization problem $\mathcal{P}_{1}$ is the coupling of $\alpha_{e,n}$ with $\mathbf{W}_{k}[n]$ in $\text{C3}$ and $\text{C4}$.~In order to overcome this difficulty, we adopt the big-M formulation. In particular, we define the new optimization variable  $\tilde{\mathbf{W}}_{k,e}[n]=\alpha_{e,n}\mathbf{W}_{k}[n]$ and add the following additional constraints to the optimization problem:
 \vspace{-2mm}
	\begin{align}
	    &\text{C}12:\tilde{\mathbf{W}}_{k,e}[n]\preceq \alpha_{e,n}~P_{\max}~\mathbf{I}_{M},\\
	    &\text{C}13:\tilde{\mathbf{W}}_{k,e}[n]\preceq \mathbf{W}_{k,e}[n],~\text{C}14:\tilde{\mathbf{W}}_{k,e}[n]\succeq \mathbf{0},\\
	    &\text{C}15:\tilde{\mathbf{W}}_{k,e}[n]\succeq \mathbf{W}_{k,e}[n]- (1-\alpha_{e,n})~P_{\max}~\mathbf{I}_{M}.
	\end{align}
Besides, we introduce a set of auxiliary optimization variables $\mu_{k}[n]$ to bound the SINR from below\cite{DISAC} 
\vspace{-2mm}
\begin{equation}\label{mu}
     0\leq\mu_k[n]\leq \frac{\text{Tr}\big(\mathbf{W}_k[n]\mathbf{H}_k[n]\big)}{\sum_{i\neq k}^{}\text{Tr}\big(\mathbf{W}_i[n]\mathbf{H}_k[n]\big)+\sigma^2_k},
\end{equation}
where $ \textbf{H}_k[n]=\textbf{h}_k[n]\textbf{h}^{H}_k[n] $.
However, \eqref{mu} is still non-convex. To overcome this issue, by introducing auxiliary variable $\phi_{k}[n]$, we can rewrite $\text{C}2$ as follows:
\vspace{-2mm}
		\begin{align}\label{mui}  &   \text{C2a}:\text{Tr}\big(\mathbf{W}_k[n]\mathbf{H}_k[n]\big)\geq \mu_k[n]\phi_{k}[n],\\
		&\text{C2b}: \sum_{i\neq k}^{}\text{Tr}\big(\mathbf{W}_i[n]\mathbf{H}_k[n]\big)+\sigma^{2}_{k}\leq\phi_{k}[n].\label{phi}
		\end{align}
		The left-hand side of \eqref{mui} is convex. However, the right-hand side is a product of two terms and not convex. Nevertheless, we can rewrite the product of the two terms as
		\begin{align}\label{26}
 \hspace*{-2mm}\mu_k[n]\phi_k[n]=\frac{1}{2}\Big[\big(\mu_k[n]+\phi_k[n]\big)^2\hspace*{-0.5mm}-\hspace*{-0.5mm}\big(\mu_k^2[n]+\phi_k^2[n]\big)\Big].
		\end{align}
Note that \eqref{26} is a difference of convex (DC) functions \cite{Ata}.~As a result, the first-order Taylor approximation can be adopted to obtain a concave function and $\mu_k[n]\phi_k[n]$ can be bounded as follows:
	\begin{align}\label{nu}
	  &\mu_k[n]\phi_k[n] \geq 0.5\big(\mu_k[n]+\phi_k[n]\big)^2-\mu_{k}^{(t)}\big(\mu_{k}[n]-\phi_{k}^{(t)}[n]\big)\nonumber\\&\hspace*{18mm}
- \phi_{k}^{(t)}[n]\big(\phi_{k}[n]-\phi_{k}^{(t)}[n]\big)\triangleq \nu_{k}[n],
	\end{align}
where ${t}$ denotes the iteration index for SCA.~Next, we relax the integer variable to a continuous one and rewrite  $\text{C11}$ as follows:
\begin{align}
    \text{C11a}: 0\leq \alpha_{e,n}\leq 1,~ 
    \text{C11b}: \sum_{e=1}^{E}\sum_{n=1}^{N}\alpha_{e,n}-\alpha^{2}_{e,n}\leq 0.
\end{align}
Constraint $\text{C11b}$ is a DC function and we use first-order Taylor approximation to convert this non-convex constraint to the following convex constraint
\begin{align}
 \overline{\text{C11b}}:\sum_{e=1}^{E}\sum_{n=1}^{N}\big(\alpha_{e,n}-\alpha^{(t)}_{e,n}(2\alpha_{e,n}-\alpha^{(t)}_{e,n})\big)\leq 0.
\end{align}
Now, we introduce a penalty factor $\tau$ to add $\overline{\text{C11b}}$ to the objective function. Thus, optimization problem $\mathcal{P}_{1}$ can be restated as follows
\vspace{-2mm}
	\begin{align}
	 \mathcal{P}_{2}: &\mathop {{\rm{min}}} \limits_{\scriptstyle{\widetilde{\mathbf{\Xi}}}}\frac{1}{N}\sum_{n=1}^{N}\bigg(\sum_{k=1}^{K} \text{Tr}(\mathbf{W}_{k}[n]) +\nonumber\\&\sum_{e=1}^{E}\alpha_{e,n}P_{\text{hover}}[n]+(1-\sum_{e=1}^{E}\alpha_{e,n})P_{\text{fly}}(\mathbf{v}[n])\bigg)+\nonumber\\&\tau\bigg(\sum_{e=1}^{E}\sum_{n=1}^{N}\big(\alpha_{e,n}-\alpha^{(t)}_{e,n}(2\alpha_{e,n}-\alpha^{(t)}_{e,n})\big)\bigg)\nonumber\\
	\text{s.t.}~~
	&\text{C}1:\sum_{k=1}^{K}\text{Tr}(\mathbf{W}_{k}[n])\leq P_{\max},\nonumber\nonumber\\
	& \text{C2c}:\frac{1}{N}\sum_{n=1}^{N}\log_2(1+\mu_{k}[n])\geq R_{\min}^{k},\nonumber\label{cc_th}\\
&\overline{\text{C2a}}:\text{Tr}\big(\mathbf{W}_k[n]\mathbf{H}_k[n]\big)\geq \nu_{k}[n],~\text{C2b},\nonumber\\
	&\text{C16}:\text{Rank}(\mathbf{W}_{k})\leq 1,~\text{C3-C8},~\text{C11a},\text{C10-C15},
	\end{align}
where $\widetilde{\mathbf{\Xi}}=\{\mathbf{W}_{k}[n],\tilde{\mathbf{W}}_{k,e}[n],\alpha_{e,n},\mu_{k}[n],\phi_{k}[n]\}$ is the new set of optimization variables. Here, penalty factor $\tau$ can be used to penalize the objective function to enforce binary values for $\alpha_{e,n}$.
Now, by dropping the rank-one constraint \text{C16} on $\mathbf{W}_{k}[n]$ and adopting SDP relaxation, problem $ \mathcal{P}_2 $ becomes a convex optimization problem and can be efficiently solved by CVX. The tightness of the SDP relaxation can be proved following similar steps as in \cite[Appendix A]{Rank}. We omit the proof here due to space constraints. 
	\subsection{Trajectory Design and Velocity Optimization}   
	Now, we tackle the design of the trajectory and velocity of the UAV for given beamforming matrices and sensing indicators. Let us first define slack variable  $s_{k}[n]=\|\textbf{q}[n]-{\textbf{d}}_k[n]\|^2+H^2$. Next, we handle the non-convexity of the data rate constraint in \text{C2}. By introducing new auxiliary optimization variables $\beta_{k}[n]$ and $\mu'_{k}[n]$, we can bound the SINR. Consequently, $\text{C2}$ is equivalently replaced by the following constraints
 \vspace{-3mm}
\begin{align}
&\widehat{\text{C2a}}: \text{Tr}\big(\mathbf{W}_k[n]\widetilde{\mathbf{H}}_k[n]\big)\geq \mu'_{k}[n]\beta_{k}[n],\label{mu'1}\\
&\widehat{\text{C2b}}:\!\!\!\sum_{i\neq k}^{}\!\text{Tr}\big(\mathbf{W}_i[n]\widetilde{\mathbf{H}}_k[n]\big)+\sigma^2_ks_{k}[n]\leq \beta_k[n]\label{mu'2}, 
\end{align}
where $ \widetilde{\textbf{H}}_k[n]= \beta_0^2\textbf{A}{(\textbf{q}[n],{\textbf{d}_{k}})}$ and $ \textbf{A}{(\textbf{q}[n],{\textbf{d}}_k)}=\textbf{a}{(\textbf{q}[n],{\textbf{d}}_k})\textbf{a}^H{(\textbf{q}[n],{\textbf{d}}_k)}$.
The right-hand side of \eqref{mu'1} is not a convex function. Similarly as in \eqref{nu}, by adopting the first-order Taylor approximation, we obtain a convex function as $\chi_k[n] \triangleq 0.5\big(\mu^{\prime}_k[n]+\beta_k[n]\big)^2-
\mu_{k}^{\prime(t)}\big(\mu^{\prime}_{k}[n]-\mu_{k}^{\prime(t)}[n]\big)- \beta_{k}^{(t)}[n]\big(\beta_{k}[n]-\beta_{k}^{(t)}[n]\big)$, where $(t)$ denotes the SCA iteration index. The left-hand side of \eqref{mu'1} is also a non-convex function in trajectory $\mathbf{q}[n]$. 
Nevertheless, we can rewrite the left-hand side of \eqref{mu'1} as
\vspace{-3mm}
\begin{align}
&\text{Tr}\big(\mathbf{W}_k[n]\widetilde{\textbf{H}}_k[n]\big)=\beta_0^2\sum_{m=1}^{M}\!\sum_{m'=1}^{M}\!\!\mathbf{W}^k_{m,m'}[n]e^{\frac{j2\pi\frac{\hat{d}}{\lambda}H (m'-m)}{\sqrt{s_k[n]}}}\nonumber\\=&\underbrace{\beta_0^2 \sum_{m=1}^{M}\mathbf{W}^k_{m,m}[n]}_{\triangleq U_k[n](\mathbf{W}_{k})}+\beta_0^2 \sum_{m=1}^{M}\sum_{m'=m+1}^{M}|\mathbf{W}^k_{m,m'}[n]|\nonumber\\&\cos \bigg(2\pi\frac{\hat{d}}{\lambda}(m'-m)\frac{H}{\sqrt{s_k[n]}}+\phi^{{W}_{k}}_{m,m'}[n]\bigg)\!\!\triangleq\!\! J_k[n](\mathbf{W}_{k},s_{k})\label{J}, 
\end{align}
where $ \mathbf{W}_{m,m'}^k[n]$, is the element in the $ m^{\text{th}} $ row and  $ m'^{{\text{th}}}$ column of $ \textbf{W}_k[n]$. Besides, $|\mathbf{W}^k_{m,m'}[n]|$ and $ \phi^{{W}_{k}}_{m,m'}[n]$ denote the magnitude and phase of $ \mathbf{W}^k_{m,m'}[n]$, respectively. 
 Note that since the right-hand side of \eqref{mu'1} is convex, we need 
	 to find an affine approximation of $J_k[n]$ to convexify the underlying optimization problem. To this end, we propose the first-order Taylor approximation as follows
		\begin{equation}
		 \tilde{J}_k[n](\mathbf{W}_{k},s_{k})\triangleq J^{(t)}_k[n](\mathbf{W}_{k},s_{k})+\nabla_{{J}_k[n]}\big(s_k[n]-s^{(t)}_k[n]\big),\label{J_tilde}
		\end{equation}
where $\nabla_{{J}_k[n]}$ is given by 
\begin{align}
& \nabla_{{J}_k[n]}=\dfrac{-2\beta_0^2\pi\hat{d}H(m'-m)}{\lambda\big(s^{(t')}_k[n]\big)^{\frac{3}{2}}} \sum_{m=1}^{M}\sum_{m'=m+1}^{M}|\mathbf{W}^k_{m,m'}[n]|\nonumber\\&\sin \bigg(2\pi\frac{\hat{d}}{\lambda}(m'-m)\frac{H}{\sqrt{s^{(t')}_k[n]}}+\phi^{{W}_{k}}_{m,m'}[n]\bigg).
\end{align}	
By substituting \eqref{J_tilde}, \eqref{mu'1} can be restated as follows
\begin{align}
\widehat{\overline{\text{C2a}}}:~{U}_k[n](\mathbf{W}_{k})+ \tilde{J}_k[n](\mathbf{W}_{k},s_{k})\geq \chi_{k}[n]. \label{40}
\end{align}
Similarly, the left-hand side of \eqref{mu'2} can be approximated by a first-order Taylor series. As a result, the inequality in \eqref{mu'2} can be restated as
	\begin{align}
	&\widehat{\overline{\text{C2b}}}:\scalemath{0.9}{\sum_{i\neq k}^{}\big(U_i[n](\mathbf{W}_{i}) + \tilde{J}_{i}[n](\mathbf{W}_{i},s_{k})\big) +\sigma^{2}_{k}{s_{k}[n]}\leq \beta_{k}[n]\label{R}}.
	\end{align}
	Finally, we deal with the non-convexity of the power consumption model when the UAV moves. To do so, we introduce the auxiliary variable $y[n]\geq 0$, such that $y^{2}[n]=\sqrt{1+\frac{\|\mathbf{v}[n]\|^4}{4 v_0^4}}-\frac{\|\mathbf{v}[n]\|^2}{2v_0^2},$ which can be rewritten as $\frac{1}{y^{2}[n]}=y^{2}[n]+\frac{\|\mathbf{v}[n]\|^2}{v_0^2}.$ Consequently, the second term in the power consumption during UAV flight can be restated as $P_i\big(y(n)-1\big)$.~Hence, the power consumption during UAV flight can be restated as  $\tilde{P}_{\text{fly}}$=$
P_o \bigg(\frac{3 \|\mathbf{v}[n]\|^2 }{\Omega^2 r^2} \bigg) + P_i\big(y(n)-1\big)+ \frac{1}{2} r_0 \rho s A_{\mathrm r} \|\mathbf{v}[n]\|^3$.~With the above manipulations, the optimization problem is recast as
\vspace{-5mm}
\begin{align}
	 \mathcal{P}_{3}: &\mathop {{\rm{min}}} \limits_{\scriptstyle{ \mathbf{q}},\mathbf{v},s_{k},y,\mu'_{k},\beta_{k}}\mathcal{F}\triangleq\frac{1}{N}\sum_{n=1}^{N}\bigg(\sum_{e=1}^{E}\alpha_{e,n}P_{\text{hover}}[n]\nonumber+\\&(1-\sum_{e=1}^{E}\alpha_{e,n})\tilde{P}_{\text{fly}}(\mathbf{v}[n])\bigg)\nonumber\\
	\text{s.t.}~~
	&\text{C17}:\scalemath{0.9}{\frac{1}{y^{2}[n]}\leq y^{2}[n]+\frac{\|\mathbf{v}[n]\|^2}{v_0^2}}, \text{C18}:\scalemath{0.9}{s_{k}[n]\geq \|\textbf{q}[n]-{\textbf{d}}_k\|^2},\nonumber\\
	&\text{C2c}:\frac{1}{N}\sum_{n=1}^{N}\log_2(1+\mu'_{k}[n])\geq R_{\min}^{k}, \forall k,\nonumber\\
    &\widehat{\overline{\text{C2a}}}, \widehat{\overline{\text{C2b}}},\text{C7-C10}.
	\end{align}
 Problem $\mathcal{P}_{3}$ is still non-convex due to non-convex constraints \text{C17} and \text{C18}. However, these constraints can be effectively handled with the SCA technique by deriving corresponding global
lower bounds at a given local point. As a result, based on the first-order Taylor approximation of the right-hand side of \text{C17}, the following global lower bound can be obtained:
$y^{2}[n]+\frac{\|\mathbf{v}[n]\|^2}{v_0^2}\geq y^{(t)2}{[n]}+\frac{\|\mathbf{v}^{(t)}[n]\|^2}{v_0^2}+2y^{(t)}{[n]}(y{[n]}-y^{(t)}{[n]})+\frac{2\mathbf{v}^{(t)}[n]}{v_{0}^{2}}(\mathbf{v}[n]-\mathbf{v}^{(t)}[n])\triangleq g(y[n],\mathbf{v}[n])$,
where $y^{(t)}[n]$ and $\mathbf{v}^{(t)}[n]$ are the values obtained in the $t$-th iteration of SCA. Besides, since $\|\textbf{q}[n]-{\textbf{d}}_k\|^2$ is a convex function with respect to $\mathbf{q}[n]$, we obtain the global lower bound based on the first-order Taylor approximation at the given point $\mathbf{q}^{(t)}[n]$ as $\|\textbf{q}[n]-{\textbf{d}}_k\|^2\geq \|\textbf{q}^{(t)}[n]-{\textbf{d}}_k\|^2+2(\textbf{q}^{(t)}[n]-{\textbf{d}}_k)^{T} (\textbf{q}[n]-\textbf{q}^{(t)}[n])\triangleq f(\textbf{q}[n],{\textbf{d}}_k)$.
This leads to the following convex optimization problem 
\begin{align}
  \mathcal{P}_{4}: &\mathop {{\rm{min}}} \limits_{\scriptstyle{ \mathbf{q}},\mathbf{v},s_{k},y,\mu'_{k},\beta_{k}}\mathcal{F}\nonumber\\
	\text{s.t.}~&\widetilde{\text{C}17}:\frac{1}{y^{2}[n]}\leq g(y[n],\mathbf{v}[n])\nonumber,\hspace{1mm}\widetilde{\text{C}18}:s_{k}[n]\geq f(\textbf{q}[n],{\textbf{d}}_k),\nonumber\\
	&\text{C2c},\widehat{\overline{\text{C2a}}},\widehat{\overline{\text{C2b}}},\text{C7}-\text{C10}.
\end{align}
In each iteration $t$, we update the solution set and  efficiently solve $\mathcal{P}_{4}$ by CVX.
\subsection{Overall Algorithm}
The proposed solution based on AO is summarized in \textbf{Algorithm 1}. Note that for sufficiently large penalty factors $\tau$ in $\mathcal{P}_{2}$, the objective function of $\mathcal{P}_{1}$ is non-increasing in each iteration of \textbf{Algorithm 1} and converges to a high-quality suboptimal solution with polynomial time computational complexity \cite{AO}. The computational complexity of \textbf{Algorithm 1} is given by $\mathcal{O}\Big(\mathrm{log}(1/\varepsilon_{\text{AO}})\big((2N+K+3NK+4EKN+2EN+2E)M^{3}+(2N+K+3NK+4EKN+2EN+2E)^{2}M^2+(4N+3NK+K+NE)(M)^3+(4N+3NK+K+NE)^{2}M^2\Big)$, where $\mathcal{O}\left ( \cdot  \right )$ is the big-O notation and $\varepsilon_{\text{AO}}$ is the convergence tolerance of \textbf{Algorithm 1}. 


\begin{figure*}[htbp]
	\centering
	\vspace{-2.9mm}
	\subfigure[UAV trajectories.]{
		\label{figure3a}
		\vspace{-3mm}
		\includegraphics[width=5.6cm]{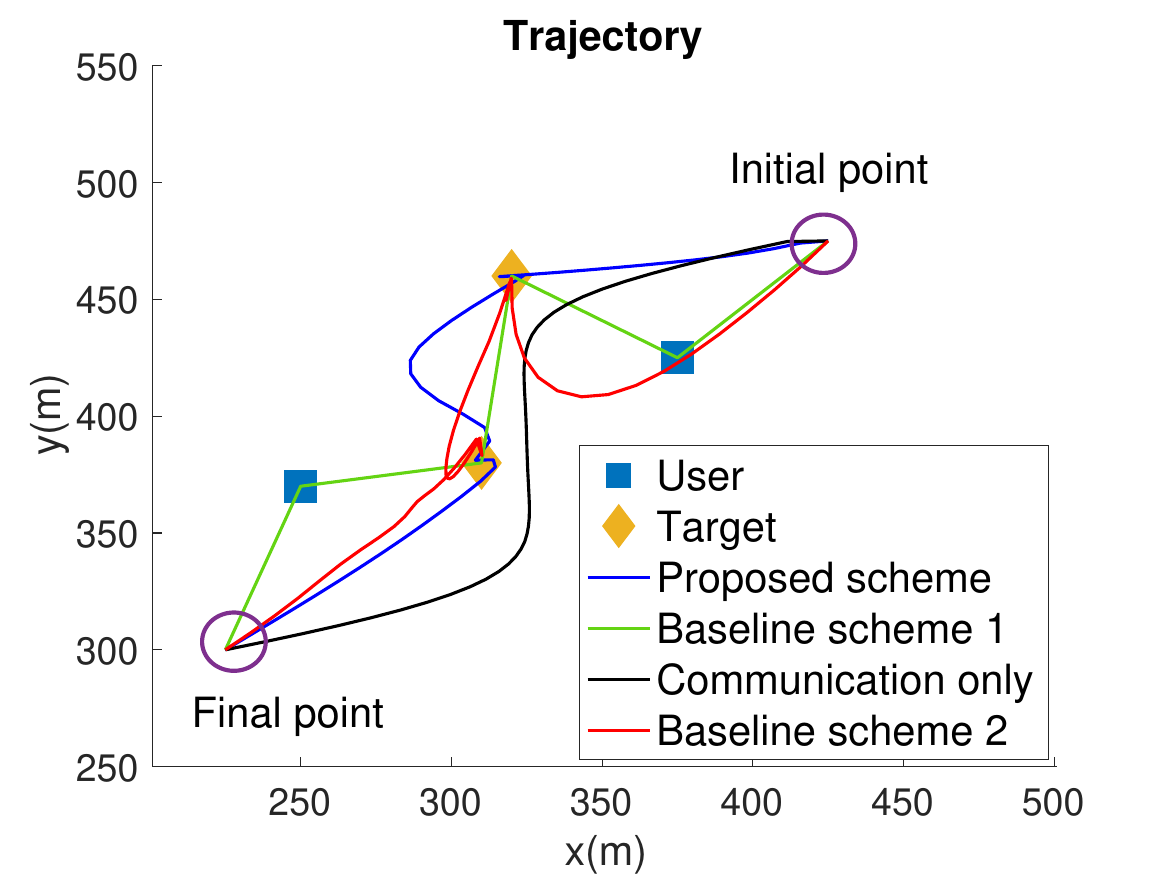}
	}
	\subfigure[Velocity of the UAV versus time (s).]{
		\label{figure3b}
		\includegraphics[width=5.6cm]{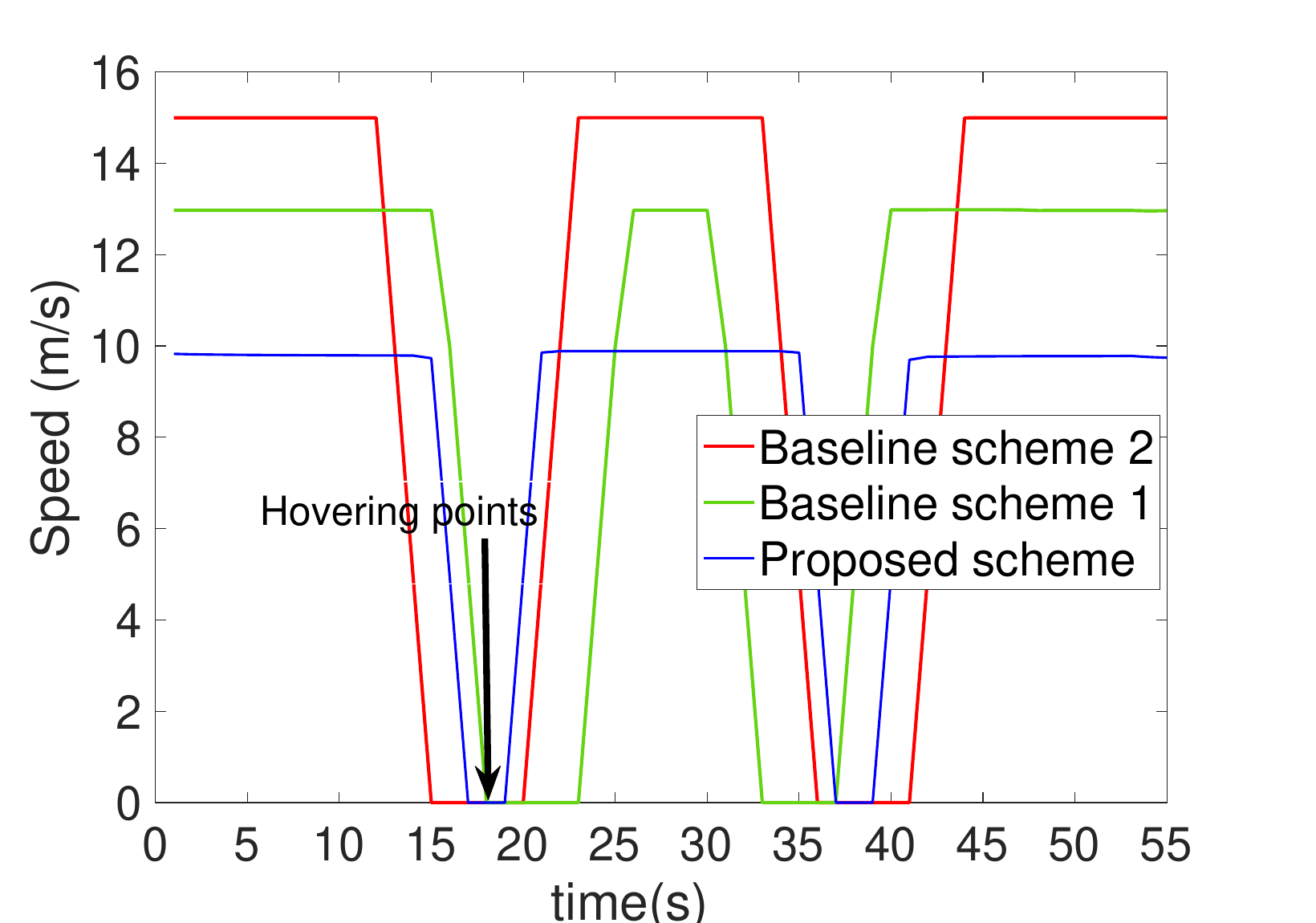}
	}
	\subfigure[Average power consumption versus the minimum required sensing SNR.]{
		\label{figure3c}
		\includegraphics[width=5.100cm]{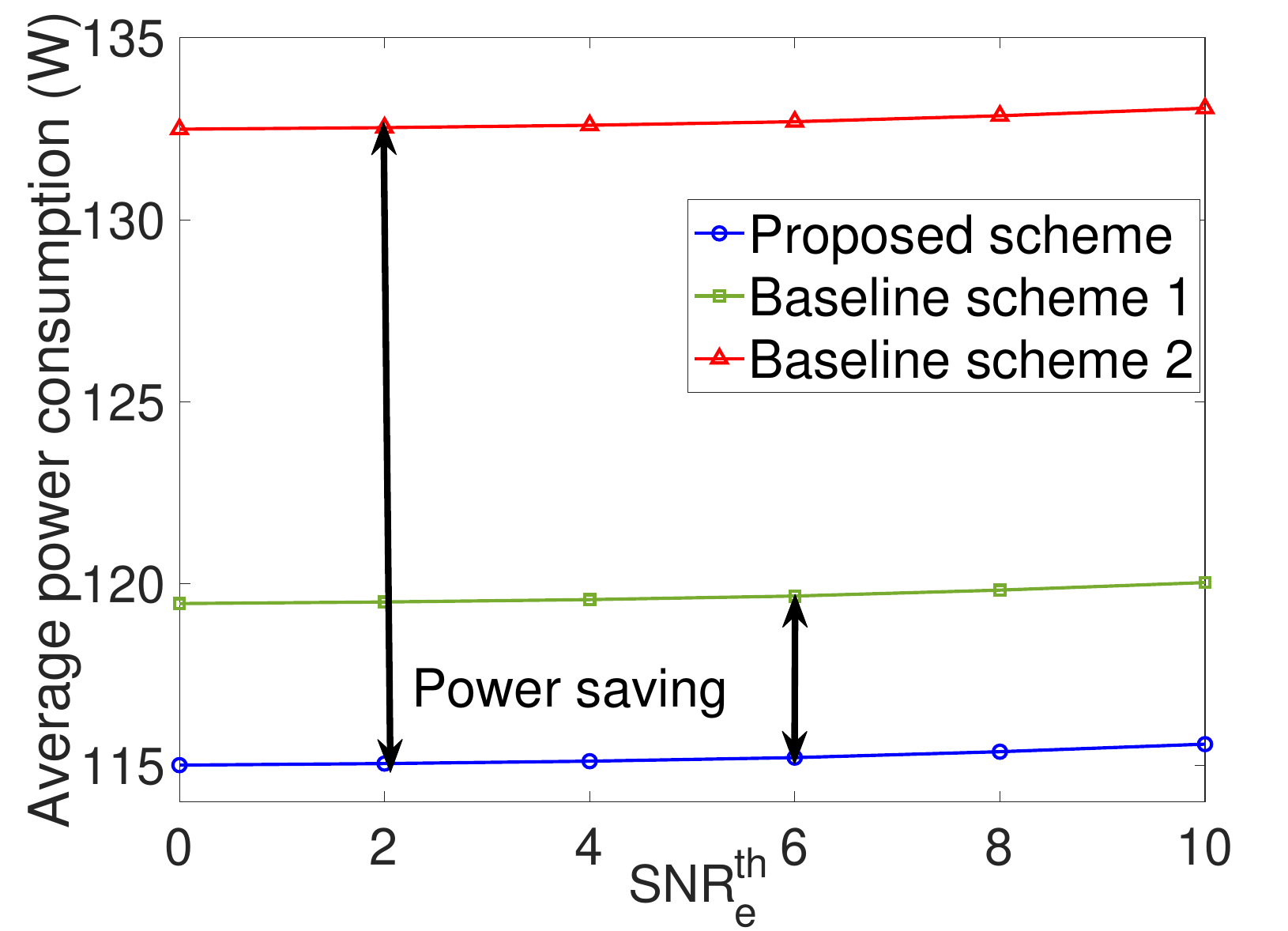}
	}
	\vspace{-3mm}
	\caption {Trajectory, velocity, and average power consumption of the UAV.}
	\label{figure6}
	\vspace{-3mm}
\end{figure*} 
\vspace{0mm}
\section{Simulation Results}
\vspace{0mm}
In this section, we evaluate the performance of the proposed algorithm via computer simulations.~We consider an area of $0.5$ km $\times$ $0.5$ km with $K=2$ communication users and $E=2$ sensing targets. The UAV is equipped with $M=6$ antennas and the minimum long-term sensing SNR at the UAV is $\text{SNR}_{e}^{\text{th}}=0$~dB \cite{ThUAVISAC}.~Moreover, the maximum flight speed of the UAV is $v_{\text{max}}=15$ m/s and the flight altitude is $H=40$ m. Besides, the channel power gain at reference distance $d_{0}=1$ m is $\beta_{0}=-30$ dB. Unless specified otherwise, we set  $\sigma^{2}_{e}=\sigma^{2}_{k}=-110$ dBm, $P_{\text{max}}=40$ dBm, $R_{\min}=1$~bps/Hz, $a_{\max}=5$~m/s$^{2}$,~$N_{s}^{\max}=5$, $T=55$~s, $D=5$~m, and $\delta_{t}=1$~s. To investigate the power saving achieved by the proposed scheme, we compare it with two baseline schemes. For baseline scheme 1, we adopt a heuristic trajectory where the UAV visits each communication user and sensing target based on the minimum distance path while optimizing the downlink information and sensing beamformers, the sensing indicator, and the velocity. For baseline scheme 2, we adopt zero-forcing beamforming for information transmission and assume an additional beam for sensing. We further assume that the velocity is fixed, i.e., $v_{\max}=15$ m/s, and omit C9. Then, we jointly optimize the sensing beam, sensing indicator, and trajectory based on a modified version of $\mathcal{P}_{1}$. 

Figs. 3(a) and 3(b) depict the trajectory and velocity of the UAV during its mission. In particular, for the proposed scheme, the UAV starts flying from the initial point towards the location of the first target while transmitting data to the communication users. During this time, the UAV also controls its velocity to minimize power consumption. Fig.~3(b) shows that, for the proposed algorithm, the UAV prefers a speed of around 10 m/s rather than the maximum speed since this speed minimizes the aerodynamic power consumption of the UAV. When approaching the first sensing target, the UAV gradually reduces its velocity to zero hover above the target for sensing. Next, the UAV flies towards the second target and senses it while hovering. 
Finally, the UAV flies towards the final point while supporting the communication users. It is interesting to observe that the trajectory of the UAV is curved. This is because in order to save power, the UAV tries to fly at the optimum velocity and as close as possible to the communication users. Fig. 3(a) also shows the trajectory of the UAV when there is no sensing requirement. In this case, in order to save power, the UAV prefers to fly  between both users to simultaneously support them.~From Fig. 3(b), we can observe that for baseline scheme 1, as the trajectory is not optimized, the UAV needs to fly with a higher velocity to complete its mission which leads to a higher transmit power consumption as can be observed in Fig. 3(c).~Another interesting observation is that the proposed algorithm leads to shorter hovering times compared to the baseline schemes, since because of the optimization of the sensing indicator, beamformers for information and sensing, trajectory, and velocity of the UAV, less time is needed to complete the sensing tasks.

Fig. 3(c) shows the average power consumption versus the sensing SNR requirement. The UAV's average power consumption for the proposed scheme and the baseline schemes is monotonically nondecreasing with respect to the minimum SNR threshold for sensing. This is because to meet more stringent sensing requirements, the UAV needs to  transmit with higher power. Moreover, we can observe the impact of the velocity and trajectory optimization on the power consumption of the UAV. In particular, the proposed scheme requires less power compared to baseline scheme 1, which employs a fixed trajectory, as the trajectory design introduces extra degrees of freedom. Moreover, baseline scheme 2 also causes a higher power consumption in comparison with the proposed scheme. In fact for baseline scheme 2, in addition to the fixed beamforming policy which leads to a higher transmit power, a considerable amount of aerodynamic power is consumed because of the fixed high UAV velocity.


\begin{algorithm}[t]
    \footnotesize
    \captionof{algorithm}{Proposed resource allocation framework.}
     \label{algorithm}
     1.\quad Initialize $\mathbf{W}_{k}^{(t)}[n]$,~$\alpha_{e,n}^{(t)}$,~$\mathbf{v}^{(t)}[n]$,~$\mathbf{q}^{(t)}[n]$ ,~$\mu_k^{(t)}$,~$\phi_k^{(t)}$, $\beta_k^{(t)}$, $\mu_k^{\prime(t)}$, $\tau \gg 1$, $t$~ \text{(iteration index)},~$\varepsilon_{\text{AO}}$.\\
     \textbf{Repeat} \\
      2.\quad Solve $\mathcal{P}_2$ for given $\mathbf{v}[n]=\mathbf{v}^{(t)}[n]$,~$\mathbf{q}[n]=\mathbf{q}^{(t)}[n]$ and obtain $\mathbf{W}_{k}^{(t+1)}[n] $, and $\alpha_{e,n}^{(t+1)}$.\\
3. \quad Solve $\mathcal{P}_4$ for given $\mathbf{W}_{k}[n]= \mathbf{W}_{k}^{(t+1)}[n]$ , $\alpha^{(t+1)}_{e,n}$, and obtain $\mathbf{v}^{(t+1)}[n]$,~$\mathbf{q}^{(t+1)}[n]$.\\
5. \quad Set~$t=t+1$\\
       6. \quad \textbf{until}~$\frac{\mathcal{O}bj^{(t)}-\mathcal{O}bj^{(t-1)}}{\mathcal{O}bj^{(t-1)}}\leq \varepsilon_{\text{AO}}$.\\ 
\end{algorithm}
 
\section{Conclusion}
 The joint resource allocation and trajectory design for a multi-user multi-target UAV-based ISAC system was studied in this paper.~We formulated the algorithm design as an optimization problem for minimization of the total UAV power consumption while taking into account the QoS requirements of the users and sensing tasks.~Specifically, for the sensing task, synthesizing a focused beam with small sidelobes, achieving a required accumulated sensing SNR, and ensuring that the UAV hovers above the target during sensing were considered. A computationally-efficient AO-based algorithm was developed for handling the resulting non-convex MINLP to obtain a high-quality suboptimal solution. Simulation results revealed dramatic power savings enabled by the proposed scheme compared to two baseline schemes.
\bibliography{Mybib}
\bibliographystyle{IEEEtran}
\end{document}